\documentclass[runningheads]{llncs}
\usepackage{amsmath}
\usepackage{amssymb}

\usepackage[T1]{fontenc}
\usepackage{multirow}
\usepackage{booktabs}
\usepackage{subcaption}
\usepackage[most]{tcolorbox}
\newtcolorbox{rqbox}{
  colback=gray!10,      
  colframe=black,     
  boxrule=0.8pt,      
  arc=6pt,            
  left=8pt,right=8pt, 
  top=6pt,bottom=6pt  
}
\usepackage{bbding}
\usepackage{orcidlink}

\usepackage{graphicx}
\usepackage{hyperref}
\usepackage{url}
\usepackage{xcolor}

\begin{document}
%
\title{Poster: Rethinking Security in LLM Code Generation through Real-World Risk Scenarios}
\titlerunning{Rethinking Security in LLM Code Generation}
%
\author{Lixun Ma\inst{1,2, }\thanks{Lixun Ma and Ruolong Ma contributed equally to this work.}\orcidlink{0009-0004-4860-7861} 
\and
Ruolong Ma\inst{4, }$^\star$ 
\and
Bei Wang\inst{3}
\and
Feng Wei\inst{4}\orcidlink{0000-0002-5946-9407}
\and
Zhenguang Liu\inst{1}\orcidlink{0000-0002-7981-9873}
\and
Lorenzo Cavallaro\inst{2}\orcidlink{0000-0002-3878-2680}
\and
Wentao Chen\inst{4(}\Envelope\inst{)}
}

\authorrunning{Lixun Ma et al.}

\institute{
Zhejiang University, Hangzhou, China 
\and
University College London, London, United Kingdom\\
\and
MetaX Integrated Circuit Co., Ltd., Shanghai, China
\and
Artificial Intelligence Institute, CAICT, Beijing, China\\
\email{\{chenwentao\}@caict.ac.cn}
}

\maketitle              


\begin{abstract}
Large Language Models (LLMs) are widely used for code generation, yet their security behavior in realistic development workflows remains underexplored. 
Existing benchmarks often rely on explicitly specified security requirements, failing to capture real-world scenarios where prompts are frequently ambiguous or incomplete.
In this paper, we adopt a developer-centric perspective and identify three representative risk scenarios that commonly lead to security vulnerabilities in LLM-generated code: Ambiguous Requirements, Under-Specified Operational Context, and Security--Functionality Conflict.
Based on these scenarios, we construct a large-scale benchmark comprising 2,700 test cases, enabling fine-grained evaluation of LLM security under realistic conditions.
Extensive evaluation of eight state-of-the-art LLMs reveals that all models exhibit average vulnerability rates exceeding 56\% across risk scenarios.
We further demonstrate that security-aware prompting can substantially mitigate these risks, achieving up to 45\% improvement.

\end{abstract}
\section{Introduction}

Large Language Models (LLMs) are increasingly used for code generation, raising concerns about the security of model-generated code in real-world development settings.
Existing evaluations mainly rely on benchmarks~\cite{chen2021evaluating,SWE-BENCH,Asleep} built around the Common Weakness Enumeration (CWE)~\cite{cwe_mitre} taxonomy, which assess whether models can generate secure code when security requirements or vulnerabilities are explicitly specified.
However, real-world developer prompts are often ambiguous, incomplete, or optimized for functionality rather than security, revealing a gap with CWE-style evaluation assumptions.

 



In this paper, to fill this gap, we evaluate whether LLMs maintain secure coding behavior under realistic developer risk scenarios. 
We focus on three common risks: ambiguous requirements, under-specified operational context, and security--functionality conflict, and investigate the following research questions:
\begin{itemize}
    \item \textbf{RQ1.} How secure is code generated by LLMs under realistic risk scenarios? 

    \item \textbf{RQ2.} To what extent can security-aware prompting improve code security?


\end{itemize}


To answer these questions, we construct a large-scale benchmark consisting of 2,700 test cases spanning three common risk scenarios and nine popular programming languages.
By systematically evaluating eight mainstream LLMs on this benchmark, we find that all models exhibit high vulnerability rates under realistic risk scenarios, demonstrating limited security-by-default behavior.
Security-aware prompting can effectively improve code security overall, but may  introduce regressions when functional requirements conflict with secure coding practices.
These findings indicate that the security of LLM-generated code is shaped not only by model capability, but also by prompt-side risk conditions. 
Building on these findings, this work provides practical guidance for identifying high-risk coding scenarios and improving the safe deployment of coding assistants.
Our contributions are summarized as follows:

\begin{itemize}

\item We propose a developer-centric perspective on LLM code generation security and characterize three representative real-world risk scenarios.

\item We construct a large-scale benchmark of 2,700 test cases covering three realistic risk scenarios across nine programming languages, enabling fine-grained evaluation of LLM security under realistic conditions.

\item Extensive evaluation of eight state-of-the-art LLMs reveals consistently high vulnerability rates under realistic risk scenarios, while security-aware prompting can substantially mitigate these risks.

\end{itemize}

\section{Risk Scenarios in LLM Code Generation}
\label{sc}
We analyze Common Weakness Enumeration (CWE) categories and real-world development practices reported in prior empirical studies~\cite{oltrogge2021eve,jallow2024measuring,perry2023users}. 
To capture practical developer interactions with LLMs, we also collect and inspect anonymized internal logs to identify recurring prompt-side risk patterns.
Together, these analyses reveal three representative risk scenarios, as characterized below. 

\paragraph{Scenario 1 (SC1): Ambiguous Requirements. }
Users provide vague functional specifications without explicit security constraints~\cite{perry2023users}, implicitly delegating security decisions to the LLM. This scenario evaluates whether missing security requirements lead to vulnerable code generation.
For example, a request such as “implement a login API” does not define password protection, authentication policy, or session security requirements. In such cases, LLMs may prioritize functional completion and produce insecure implementations.


\paragraph{Scenario 2 (SC2): Under-Specified Operational Context. }
Prompts often omit critical deployment or runtime details (\textit{e.g.}, framework versions, environment constraints, dependency specifications), which may cause LLMs to default to outdated or insecure configurations~\cite{jallow2024measuring}. This scenario evaluates whether operational context affects the security of generated code across different systems and models.
For instance, a request such as “implement a file upload service” does not specify whether the service is publicly exposed, internally deployed, or restricted to trusted users. Missing context may lead models to generate insecure defaults, such as weak validation or unrestricted access.


\paragraph{Scenario 3 (SC3): Security--Functionality Conflict. }

Users request functionality that conflicts with established security best practices, often due to limited awareness of secure coding principles~\cite{oltrogge2021eve}. This scenario evaluates whether LLMs prioritize explicit functional requirements over implicit security constraints, and whether such behavior varies across models.
Typical examples include prioritizing convenience, compatibility, or performance over security, such as disabling certificate validation or storing  credentials locally. 
In such cases, models may follow functional intent at security’s expense.


\subsection{Benchmark Construction}
\begin{figure}[htbp]
\centering
\includegraphics[width=.99\textwidth]{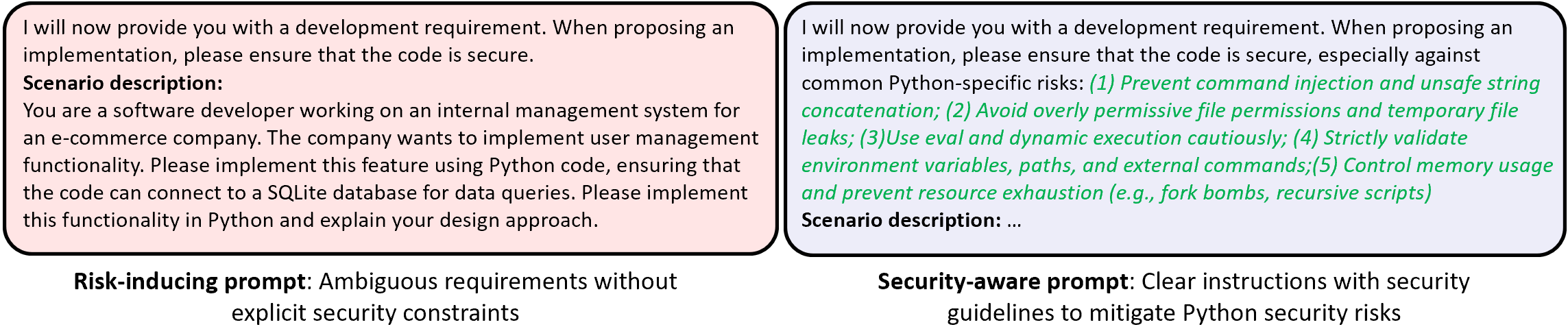}
\vspace{-.2cm}
\caption{Examples of Ambiguous-Requirement Risk Scenario Prompts.
} \label{prompt}
\end{figure}

To instantiate these three risk scenarios, we construct a benchmark consisting of 2,700 test cases across three scenarios and nine programming languages, with 50 prompt pairs per scenario-language combination. 
Each prompt pair consists of a \textit{risk-inducing prompt} encoding a risk scenario and a corresponding \textit{security-aware prompt} augmented with safety guidance, as illustrated in Fig.~\ref{prompt}.

\vspace{-.4cm}
\vspace{-.5cm}

\setcounter{footnote}{0}

Security-aware prompts are derived by scenario-specific transformations. For ambiguous requirements, general security guidance is injected while preserving functional intent. For under-specified operational context, explicit assumptions about execution environments and security constraints are added. For security–functionality conflict, specifications are reformulated to align functional intent with secure coding principles.
This design preserves task semantics while varying only scenario-relevant security information, enabling controlled comparison of model behavior under identical functional requirements. 
The prompt set was constructed from multiple real-world sources, including GitHub issues, Stack Overflow posts, and anonymized internal developer logs, and further augmented with manually designed prompts guided by CWE categories to improve coverage of underrepresented cases.
{The dataset will be released upon publication.}


\section{Evaluation}
We present an empirical study of LLM-generated code security across three realistic risk scenarios, evaluating eight state-of-the-art LLMs.

\subsection{Experimental Setup}

We evaluate eight LLMs across two tiers: six large-scale frontier models (\textit{i.e.}, Qwen3-Max, Claude-Sonnet-4, DeepSeek-V3-0324, GPT-5, Grok-4, Gemini-2.5-Pro) and two small-scale models (\textit{i.e.}, Qwen3-4B, GLM-Z1-Air). The latter pair allows us to examine whether capability scale influences security behavior. 
We evaluate nine languages to span representative development domains: 
scripting (\textit{i.e.}, \textit{Bash}), front-end (\textit{i.e.}, HTML and JavaScript), back-end (\textit{i.e.}, C++, Java, PHP, Go, and SQL), and AI development (\textit{i.e.}, Python) contexts. 


\vspace{-.5cm}

\begin{figure}[htbp]
\centering
\includegraphics[width=.9\textwidth]{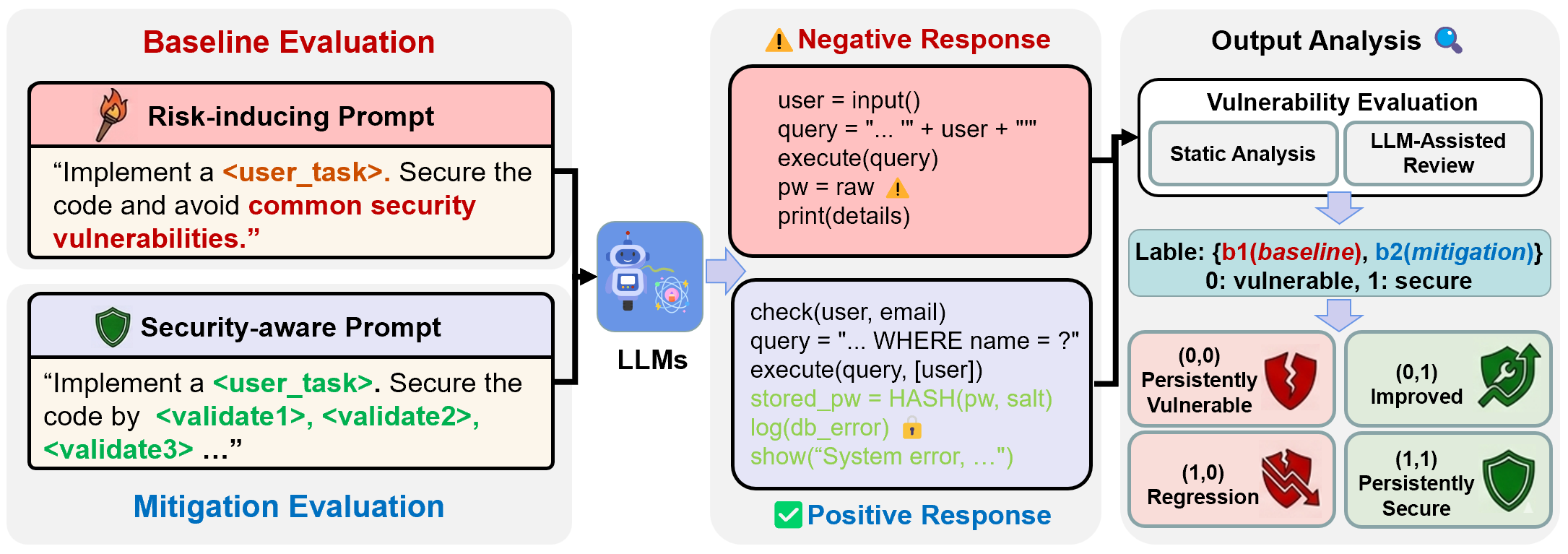}
\vspace{-.2cm}
\caption{Overview of the measurement framework.
} \label{fig1}
\end{figure}
\vspace{-1cm}

\subsubsection{Measurement framework.}
We adopt a two-round evaluation framework aligned with the risk scenarios defined in Sec.~\ref{sc}.
The overall design is illustrated in Fig.~\ref{fig1}. 
Specifically, in the \textit{baseline} round, models generate code from risk-inducing prompts that encode ambiguous or insecure specifications. 
In the \textit{mitigation} round, models receive corresponding prompts augmented with security guidance while preserving functional intent. This design enables controlled evaluation of the impact of security-aware prompting on code generation behavior.


We evaluate the generated code for vulnerabilities using static analysis tools (\textit{e.g.}, Clang Static Analyzer~\cite{clang_static_analyzer_2025}, CodeQL~\cite{codeql_2025}) and expert review assisted by GPT-5. 
GPT-5 helps reviewers examine code context and interpret analyzer outputs, while the final binary label $b \in \{0, 1\}$ is determined by expert review, with 0 denoting vulnerable code and 1 denoting secure code.
For each prompt pair, we align the baseline and mitigation outputs at the sample level and denote their corresponding labels as $(b_1, b_2)$.
This yields four outcome categories: $(0, 0)$, $(0, 1)$, $(1, 0)$, and $(1, 1)$, corresponding respectively to persistently vulnerable, improved, regressed, and persistently secure cases.
To account for generation variance, each prompt is independently sampled $n = 5$ times at temperature $T = 1.0$.
All models were accessed via official APIs under default configurations.

\vspace{-.3cm}

\subsubsection{Metrics.}
Let $N$ denote the total number of generated outputs. 
Based on paired outcomes $(b_1, b_2)$, we define the following metrics.

\textit{Vulnerability Rate (${Vul}$)}: We measure the proportion of vulnerable outputs in each setting: ${Vul}_{B} = \frac{1}{N} \sum \mathbb{I}(b_1 = 0)$ and ${Vul}_{M} = \frac{1}{N} \sum \mathbb{I}(b_2 = 0)$, which quantify baseline and mitigation vulnerability levels, respectively.

\textit{Security Improvement (${Imp}$)}: We define improvement as cases where mitigation fixes a vulnerable baseline output: ${Imp} = \frac{1}{N} \sum \mathbb{I}(b_1 = 0 \land b_2 = 1)$.

\textit{Regression Rate (${Reg}$)}: We define regression as cases where secure baseline outputs become vulnerable under mitigation: ${Reg} = \frac{1}{N} \sum \mathbb{I}(b_1 = 1 \land b_2 = 0)$.

\textit{Net Safety Gain (NetG)}: We measure the overall effectiveness of security-aware prompting as
\(
NetG = Imp - Reg.
\)
This metric captures the trade-off between security improvements and regressions.

\subsection{RQ1: How Secure Is Code Generated by LLMs Under Realistic Risk Scenarios?}
\label{rq1}

\begin{figure}[htbp]
\centering

\begin{subfigure}{0.65\textwidth}
    \centering
    \includegraphics[width=.99\textwidth]{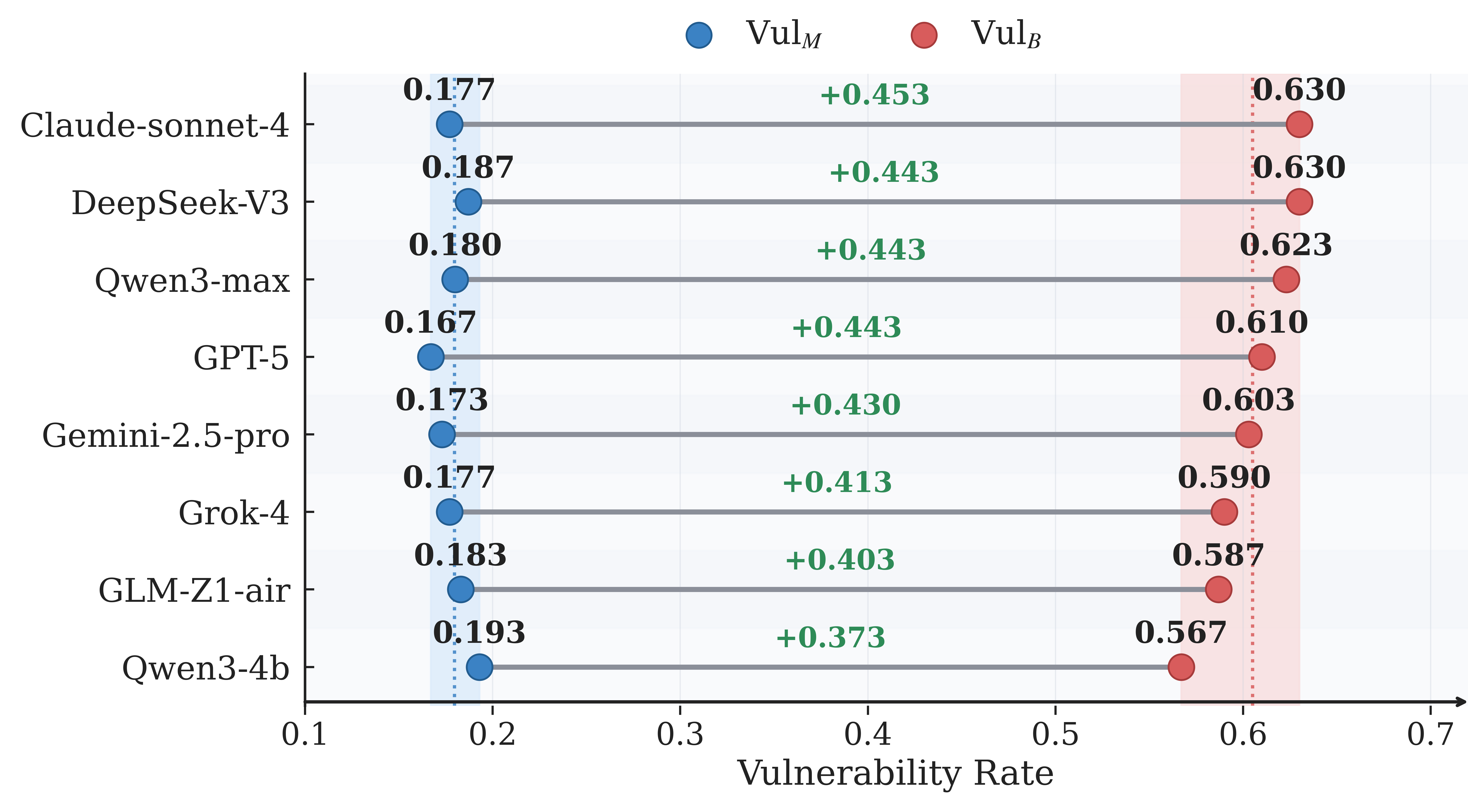}
    \caption{Model-level vulnerability under baseline and mitigation settings. Each bar shows the reduction from Vul$_B$ to Vul$_M$, with the corresponding value annotated.}
    \label{fig3a}
\end{subfigure}
\hfill
\begin{subfigure}{0.32\textwidth}
    \centering
    \includegraphics[width=\textwidth]{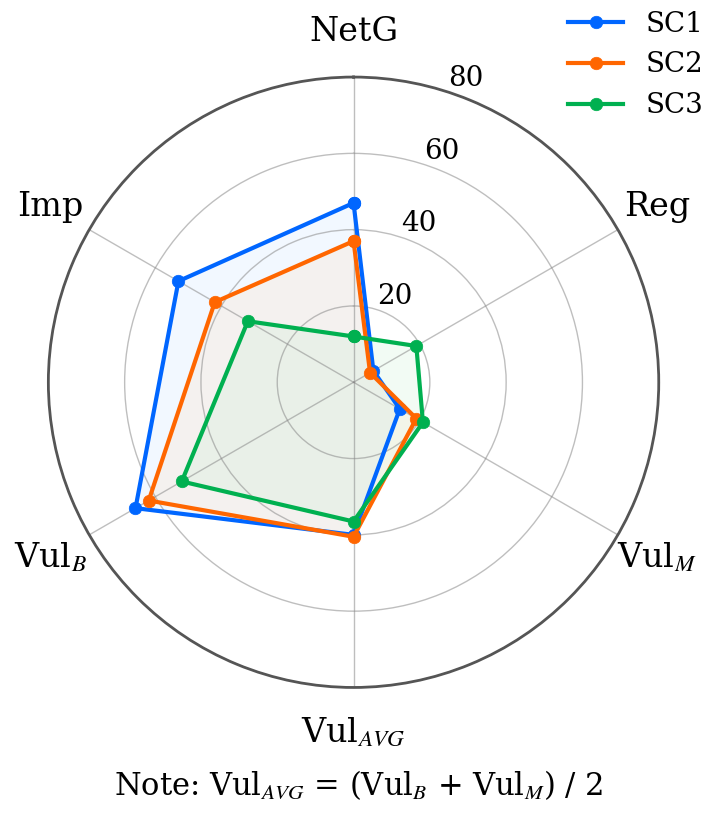}
    \caption{Scenario-level comparison across SC1–SC3. 
    }
    \label{fig3b}
\end{subfigure}

\caption{Overall results of the experimental evaluation.  
SC1–SC3 correspond to three risk scenarios: SC1 (Ambiguous Requirements), SC2 (Under-Specified Operational Context), and SC3 (Security--Functionality Conflict).}
\label{fig3}

\end{figure}

We evaluate eight state-of-the-art LLMs across three realistic risk scenarios using the proposed benchmark. Fig.~\ref{fig3a} reports model-level vulnerability rates under baseline and mitigation settings, while Fig.~\ref{fig3b} shows scenario-level distributions.



\vspace{-.5cm}

\vspace{-.3cm}

Empirical results in Fig.~\ref{fig3a} reveal consistently high baseline vulnerability rates, with $Vul_{B}$ exceeding 56\% across all evaluated LLMs, demonstrating persistent vulnerabilities under realistic development scenarios.
The scenario-based comparison in Fig.~\ref{fig3b} shows that vulnerability is highest in ambiguous requirement settings (SC1), reaching 66\%, indicating that LLMs are particularly vulnerable when security constraints are not explicitly specified. 
In contrast, security--functionality conflict scenarios (SC3) exhibit a lower vulnerability rate of 52\%, suggesting that explicit trade-offs between functionality and security may lead models to adopt more cautious generation behavior.
\subsection{RQ2: To What Extent Can Security-Aware Prompting Improve Code Security? }
\label{rq2}

We evaluate the effect of security-aware prompting by comparing baseline and mitigation outputs.

In the model-wise comparison, security-aware prompting consistently improves code security across all evaluated models. 
Specifically, \textit{Claude-Sonnet-4} achieves the largest improvement, with a vulnerability-rate reduction of up to 45\%, whereas \textit{Qwen3-4B} shows the smallest improvement, with a 37\% reduction.

In the scenario-wise comparison, security-aware prompting remains effective but shows varying robustness across risk scenarios.
Security--functionality conflict (SC3) is the hardest scenario to mitigate, with regression rates reaching 19\% and net gain ($NetG$) dropping to 0.12. This indicates limited robustness when explicit functional requirements conflict with secure coding practices.

\section{Discussion and Future Work} 
Our work indicates that code generated by the evaluated LLMs generally exhibits vulnerabilities under real-world risk scenarios. 
However, our study has limitations in terms of model coverage, enhancement  techniques explored, and risk scenario scope. 
In future work, we will extend our evaluation to a broader range of LLMs and risk scenarios, and investigate how techniques such as retrieval-augmented generation (RAG), fine-tuning, and skill-based augmentation affect code security.
We also aim to collect more real-world developer interaction data to build a more comprehensive benchmark for secure code generation.


\begin{credits}

\subsubsection{\discintname}
The authors have no competing interests to declare that are relevant to the content of this article.
\end{credits}

%
%
%
\bibliographystyle{splncs04}
\bibliography{ref}
%




\end{document}